# Ultrafast reflectivity modulation in Al$_x$Ga$_{1-x}$As/In$_y$Al$_x$Ga$_{1-x-y}$As multiple quantum well photonic crystal waveguides

A. Z. Garcia-Déniz, P. Murzyn, A. M. Fox,* D. O. Kundys, J.-P. R. Wells, D. M. Whittaker, and M. S. Skolnick

*Department of Physics and Astronomy, University of Sheffield, Sheffield S3 7RH, United Kingdom*

T. F. Krauss

*School of Physics and Astronomy, University of St. Andrews, St. Andrews KY16 9SS, United Kingdom*

J. S. Roberts

*Department of Electronic and Electrical Engineering, University of Sheffield, Mappin Street, Sheffield S1 3JD, United Kingdom*

We report an ultrafast optical tuning of the reflectivity of Al$_x$Ga$_{1-x}$As/In$_y$Al$_x$Ga$_{1-x-y}$As multiple quantum well photonic crystal waveguides using a reflection geometry, pump-probe technique. Nonlinear shifts of the photonic resonances between 800 and 900 nm are demonstrated with a measured response time of ~300–400 fs. The nonlinear shift is attributed to virtual carriers, with the response time determined by the pump pulse duration. This demonstrates ultrafast photonic switching in a high-contrast photonic structure based on III-V quantum wells.

## I. INTRODUCTION

Photonic crystals have attracted considerable interest in recent years on account of their unique properties which give new possibilities for the control of the propagation of light. The nonlinear properties of photonic crystals (PCs) are particularly attractive because of their potential use in ultrafast all-optical switches for telecommunication systems.[1] The possibility of fast optical nonlinearities has been predicted theoretically[2–5] and demonstrated experimentally in several reports,[6–11] including earlier work within our own group.[12–16] Despite the significant work carried out to date, the nonlinear properties of PCs are still not fully understood, and there is considerable scope for improved control of the optical properties, especially as regards the response time.

An ideal switch should have an instantaneous (i.e., excitation-limited) response. In an all-optical switch, this requires a response time limited only by the pump pulse duration. For ultrafast (i.e., subpicosecond) switching, a virtual carrier nonlinearity must be employed, for instance, the optical Kerr effect or the ac Stark effect. The former relies on the nonlinear refractive index, while the latter exploits the blueshift of the exciton under near-resonant pumping. For both types of effect, the incorporation of multiple quantum wells within the photonic crystal offers the possibility of enhanced nonlinearities, together with the additional possibility of polaritonic coupling of excitons to photonic modes. Despite this, there has been only a very limited amount of work on ultrafast switching in high-refractive-index PCs reported in the literature to date. The ac Stark effect has been demonstrated in an inorganic-organic system,[8] but the organic materials used are potentially susceptible to photodegradation. With the more stable purely inorganic systems, the work has so far been restricted to demonstrating the optical Kerr effect in bulk semiconductors, namely, silicon[9] and GaAs.[17]

In this paper, we demonstrate a subpicosecond reflectivity modulation in Al$_x$Ga$_{1-x}$As/In$_y$Al$_x$Ga$_{1-x-y}$As quantum well photonic crystal waveguides under both near-resonant and nonresonant pumping conditions. This is a significant development on our previous work, in which the response time was determined by the slower recombination dynamics of real carriers,[12–16] and also on ultrafast work by other groups, which has so far been restricted to bulk materials.[9,17] We have used the same time-resolved surface reflectivity pump-probe technique as previously, but we now report an ultrafast response time that we attribute to virtual carrier effects. The switching times obtained in this way are of the order of 300–400 fs and are only limited by the duration of the pump pulses used in our experiment.

## II. SAMPLE AND EXPERIMENTAL DETAILS

The base for all the photonic structures studied here was the same planar waveguide grown by MOVPE (metalorganic vapor phase epitaxy). The core was 400 nm thick and consisted of Al$_{0.2}$Ga$_{0.8}$As with an embedded active region containing five periods of 9.6-nm-thick In$_{0.12}$Al$_{0.2}$Ga$_{0.68}$As quantum wells with 10-nm-thick Al$_{0.2}$Ga$_{0.8}$As barriers. The waveguide was realized with the air and a thin 10 nm GaAs capping layer as the top cladding and 1500-nm-thick Al$_{0.6}$Ga$_{0.4}$As layer as the bottom cladding. Photonic crystals were patterned onto the wafer from the top surface by electron-beam lithography and chemically assisted ion beam etching. Numerous one-dimensional (1D) photonic structures with different periods and air fill factors [see, e.g., Fig. 1(a)] were patterned onto the same wafer. Each photonic structure was a square with dimensions 80×80 $\mu$m². The periods of the structures ranged from 330 to 880 nm with air fill factors of 10%, 15%, 20%, and 30%. The etch depth of 850 nm ensured that the structure pattern went well through the core, reaching into the bottom cladding. Room temperature photoluminescence (PL) measurements of the structure show exciton emission at 799 nm (see Fig. 2).

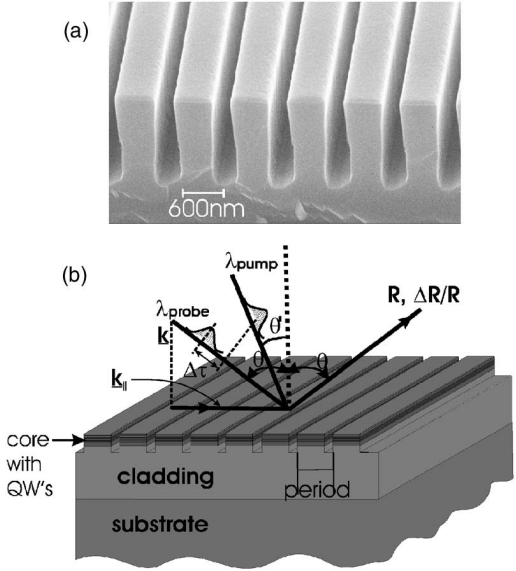

FIG. 1. (a) Scanning electron microscope picture of one of our 1D photonic crystals. (b) Schematic diagram of the 1D photonic crystal waveguides and the experimental geometry used in this work.

The measurements were carried out at room temperature by a time-resolved pump-probe technique in reflection geometry. One of the beams was generated with an optical parametric amplifier (OPA), which provided continuous tunability from 200 nm to 18 $\mu$m. The OPA was pumped by a regeneratively amplified, mode-locked, Ti:sapphire (Ti:S) laser with a pulse length of 150 fs and 1 kHz repetition rate. A small portion of the Ti:S beam, which could be tuned from 760 to 840 nm, was used as a second beam in our pump-probe experiment.

To probe the reflectivity spectra of our samples we have used a white-light continuum, which was generated by focusing one of the beams into a 1-mm-thick sapphire plate. The collimated white light probe beam was incident on the sample and was imaged onto a pinhole after being reflected, to spatially filter the signal from the individual photonic crystal sample. Figure 1(b) shows the experimental geometry. The reflected spectrum along with the reference white-light beam was focused into a spectrometer and recorded with a charge-coupled device camera. The reference beam allowed the normalized reflectivity spectrum to be obtained to compensate for any intensity fluctuations of the white light probe. The geometry of the experiment did not allow for the observation of the full photonic band. However, a significant amount of light could still be coupled into the photonic crystal, resulting in photonic resonances (coupling features) present in the reflectivity spectra. The incident light has to satisfy the phase-matching condition $k_\parallel = (\omega/c)\sin\theta$, where $k_\parallel$ is the in-plane wave vector, $\omega$ is the angular frequency, and $\theta$ is the coupling angle. The strength and the width of a given resonance depend on the coupling constant and the $k$ vector. The resonant modes investigated here were situated between 800 and 900 nm. The pump beam was focused onto the sample with a spot size of 100–200 $\mu$m depending on the pump beam wavelength and was incident on the sample at $\sim 15°$ relative to the probe beam.

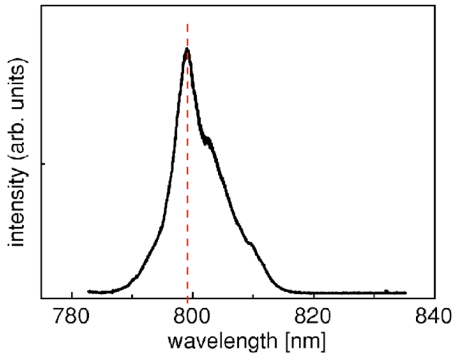

FIG. 2. (Color online) Room temperature excitonic PL spectrum of the multiple quantum well.

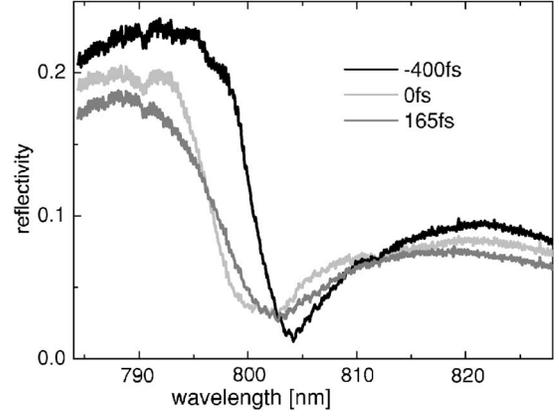

FIG. 3. Reflectivity spectra for near-zero time delay showing a blueshift of the photonic resonance corresponding to the ac Stark effect. The pump intensity was 6 GW/cm$^2$ at 803 nm.

## III. EXPERIMENTAL RESULTS

### A. Near-resonant pumping

We first consider the case of near-resonant pumping, for which we selected a photonic structure with a coupling resonance at 804 nm (i.e., in the vicinity of the exciton energy as shown in Fig. 2). The white-light continuum probe was generated using 1.3 $\mu$m pulses from the OPA laser. The Ti:S beam, which could be tuned in a limited range, i.e., 790–810 nm, without the need for OPA realignment, was then used as the pump beam.

Selected reflectivity spectra for one of the PCs (i.e., with 860 nm period and 30% air fill factor) are shown in Fig. 3. A blueshift of the photonic mode is clearly observed at zero delay, followed by a reduction of the blueshift and broadening of the feature at a delay of 165 fs. The experiment indicates large modulation of the reflectivity, for example, from 18% for pump off to 4% for pump on at 799 nm.

The time scale of the reflectivity modulation can be more clearly seen when differential reflectivity spectra are used for the data presentation. Figure 4 shows a set of differential reflectivity spectra vs time for the pump wavelength at

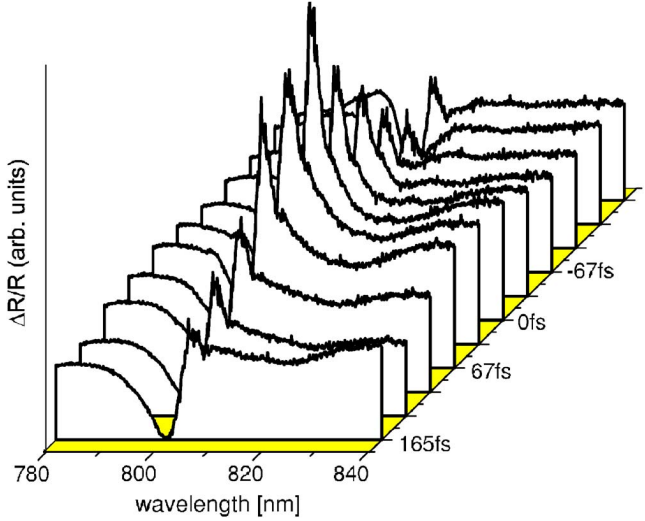

FIG. 4. (Color online) Differential reflectivity spectra for different time delays with a pump wavelength of 800 nm and intensity 6 GW/cm$^2$.

800 nm. A clear reflectivity change can be seen around 804 nm corresponding to the wavelength of the photonic resonance. A very fast modulation of the reflectivity occurs within ~350 fs, which is consistent with the cross correlation of the pump and probe beams. The duration of the reflectivity change is resolution limited and recovers very quickly after the temporal overlap of the pump and the probe beams ceases. This is shown in Fig. 5(a) where the change of the reflectivity at 804 nm (the photonic mode wavelength) is plotted vs the time delay. A pulsewidth-limited reflectivity change is observed at zero delay, followed by a rising background. We shall argue below in Sec. IV that the fast response at zero delay is a virtual carrier effect, while the rising edge is caused by the relaxation of hot carriers generated by two-photon absorption.

The wavelength dependence of the shift of the photonic mode at zero delay is plotted in Fig. 5(b). We observe that the nonlinear shift increases as the pump is tuned toward resonance with the quantum well exciton. By contrast, the reflectivity change observed for photonic modes further from the exciton energy (i.e., toward the red) was much smaller, and vanished for resonances above 830 nm. Figure 5(c) plots the magnitude of the reflectivity change against pump intensity for a pump wavelength of 800 nm. The reflectivity change is observed to vary linearly with the pump intensity, to within experimental error. When the pump wavelength falls below 800 nm, a large number of real carriers are created by single-photon absorption, and the ultrafast response is obscured.

The measured blueshift of the resonance can be used to determine the nonlinear refractive index $n_2$ from[13]

$$\frac{\Delta n}{n_0} = \frac{n_2 I}{n_0} \approx \frac{\Delta\lambda}{\lambda}, \quad (1)$$

where $\Delta n$ is the refractive index change induced by the pump, $n_0$ is the refractive index at $I=0$, $\Delta\lambda$ is the shift of the resonance, $\lambda$ is the original wavelength of the resonance, and $I$ is the intensity of the pump. The Kerr coefficient we obtain from our data is $n_2=(2.2\pm 1)\times 10^{-12}$ W$^{-1}$ cm$^2$ (i.e., close to experimental data in Ref. 19).

### B. Nonresonant pumping

We now consider the case where the pump wavelength is tuned to be far from resonance from both the photonic mode and the single-photon absorption states of the waveguide core. In our previous work we have demonstrated nonlinear reflectivity changes caused by the generation of free carriers by two-photon absorption.[12,13] In this section we shall describe results that demonstrate another effect with a much faster time response.

Pump-probe measurements were carried out for a number of samples with photonic resonances between 830 and 900 nm. The white-light probe was generated this time at 800 nm with the Ti:S beam, and the OPA was used as the pump, allowing for pump tunability over a wide spectral range. For many pumping wavelengths it was found to be difficult to observe an ultrafast reflectivity change on account of the much larger effect of the real carriers generated by multiphoton absorption. The best results were obtained for a pump wavelength of ~2.1 $\mu$m which lies below the two-photon absorption threshold.

Figure 6 shows a representative set of differential reflectivity spectra (i.e., $\Delta R/R$) as a function of time delay between the pump and probe pulses for a pump wavelength of 2.1 $\mu$m. The data shown are for the sample with the period of 860 nm and an air fill factor of 30%. The photonic resonance studied was around 865 nm. The reflectivity at the wavelength corresponding to the resonance is observed to change on a time scale of several hundred femtoseconds. At times near zero delay, the shape of the differential reflectivity spectrum indicates a bleaching and broadening of the photonic mode, with hardly any shift. At longer positive time delays, the shape changes to dispersive, which indicates that the resonance has shifted.

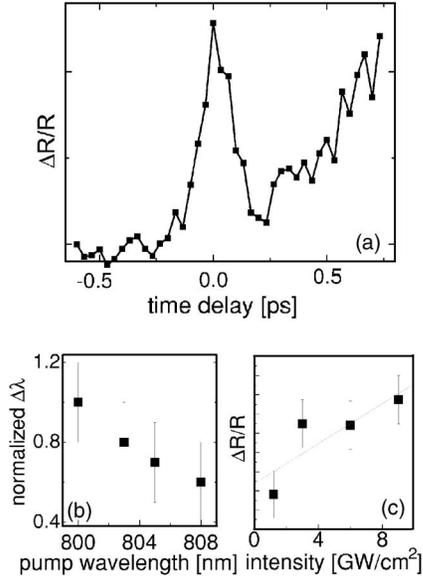

FIG. 5. (a) Differential reflectivity for a fixed wavelength of 804 nm corresponding to a photonic mode wavelength plotted as a function of the time delay (for the data from Fig. 4). (b) Magnitude of the mode wavelength shift for different pump wavelengths at a fixed intensity of 6 GW/cm$^2$. (c) Magnitude of the reflectivity change for different pump intensities for a pump wavelength at 800 nm.

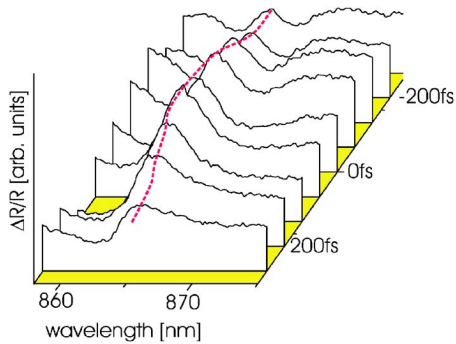

FIG. 6. (Color online) Differential reflectivity spectra vs time delays for a pump wavelength of 2.1 μm and an intensity of 1.8 GW/cm². The dotted line marks the center wavelength of the photonic resonance in the linear reflectivity spectrum.

Figure 7 shows the differential reflectivity change vs delay time at 865 nm. This reveals a time response with a symmetrical shape and a pulse-limited decay time. The full width at half maximum of the time response of ~400 fs is in good agreement with the cross correlation of the pump and the probe pulses. A slight difference in the response time compared to the near-resonant data shown in Fig. 5(a) possibly arises from the different lasers used to generate the white-light probe in the two cases. The instantaneous reflectivity modulation by IR pumping is consistent with a virtual carrier effect, and has been observed in a number of samples. The observed change of the reflectivity is approximately linear with the pump power (see inset in Fig. 7). In contrast to the near-resonant results, the response was insensitive to the wavelength of the probed photonic resonance (i.e., in the range of 830–900 nm), which is in agreement with the expectations for a nonresonant effect.

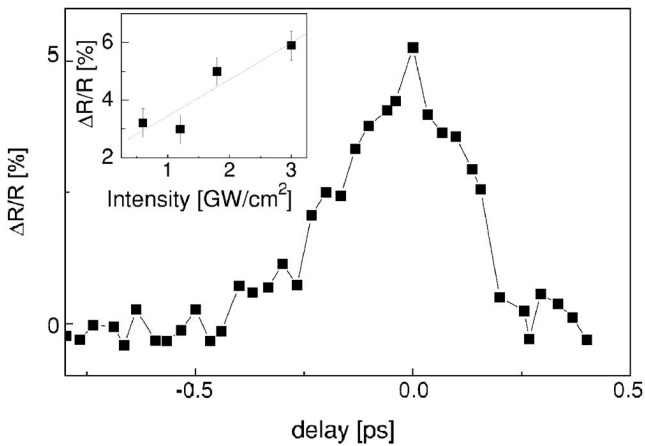

FIG. 7. Differential reflectivity at 865 nm as a function of the time delay (for the data from Fig. 6). The inset shows the magnitude of the reflectivity change for different pump powers.

## IV. DISCUSSION

In our previous work we have investigated the nonlinear shifts of the photonic coupling resonances induced by free carriers.[12–16] The free carriers can be excited by single-photon absorption when the pump photon energy is above the band gap of the waveguide core, or by multiphoton absorption for smaller photon energies. These real-carrier effects occur over a very broad range of pumping conditions, making it somewhat challenging to observe the faster nonlinearities associated with virtual carriers. Careful analysis of the experimental data is therefore required to determine the different processes contributing to the nonlinear response.

Let us first consider the case of near-resonant pumping described in Sec. III A. In these results we are considering a situation in which both the pump photon energy and the photonic mode are just below the exciton. The pump photons can produce real carriers by two-photon absorption, and can also induce virtual carrier nonlinearities. There is also the possibility that the high-energy tail of the pump pulse can excite excitons directly by single-photon absorption. These three processes can be distinguished by their different time signatures. Neither of the two real-carrier effects can adequately explain the pulse-width-limited signal near zero delay in Fig. 5(a). In the case of real carriers excited by two-photon absorption, we expect a finite rise time as the hot carriers relax to the bottom of the bands.[16] This process is clearly observed as the rising edge for positive time delays in Fig. 5(a). On the other hand, for the case of resonant excitation of excitons, we would expect a strong signal during the ~100 fs lifetime of the excitons, followed by a signal of approximately half the magnitude produced by the free carriers generated by exciton ionization.[18] We would therefore not expect the nonlinear signal to recover within the pulse duration. It is thus highly likely that the fast response observed at zero delay is a virtual-carrier effect. Given the resonant excitation conditions, the strongest virtual carrier nonlinearity is expected to be the ac Stark effect.

Further insight into the fast nonlinearity observed with near-resonant pumping comes from the wavelength and power dependence shown in Figs. 5(b) and 5(c). The exciton renormalization in the ac Stark effect is expected to change with the pump intensity and wavelength according to[8]

$$\delta\omega_{ex} \propto -\frac{|\mu E_p|^2}{(\omega_p - \omega_{ex})}, \quad (2)$$

where $\mu$ is the dipole moment, $E_p$ is the pump field, $\omega_p$ is the pump frequency, and $\omega_{ex}$ is the exciton frequency. This implies a reciprocal dependence on the detuning, and a linear dependence on the pump intensity, both of which are observed in the data, to within experimental error. The smaller nonlinear shift observed for photonic modes with larger detunings from the exciton is further evidence of the excitonic nature of the nonlinearity.

It is thus reasonably clear that the fast nonlinearity observed under near-resonant pumping can be attributed to the ac Stark effect of the quantum well excitons. We have shown that this effect is capable of producing large (i.e., >10%) absolute changes of the reflectivity on femtosecond time scales. However, in the results presented here, the effect is partially masked by the stronger nonlinearities associated with unavoidable real-carrier generation. Since these real carriers have slower response times in the picosecond range, the usefulness of the present design is somewhat limited. In practical devices, it will be necessary to improve the contrast between the virtual- and real-carrier nonlinearities. One possible way to do this would be to include more quantum wells within the core.

We now turn to consider the results for nonresonant pumping presented in Sec. III B. In this case neither the pump pulse nor the photonic mode is resonant with the quantum well exciton or the waveguide band edge. The presence of the quantum wells is therefore of little significance, and we mainly expect to observe effects relating to the nonlinearity of the much thicker $Al_xGa_{1-x}As$ core. We thus expect free-carrier nonlinearities associated with multiphoton absorption, as observed previously,[12,13] and virtual-carrier effects due to the nonresonant Kerr effect of the $Al_xGa_{1-x}As$. In the former case, we expect a finite response time due to the lifetime of the carriers, while in the latter we expect a pulse-width-limited response. Neither of these effects is expected to be particularly sensitive to the mode wavelength, as was indeed the case for the data presented in Sec. III B.

The main result for the nonresonant conditions is presented in Fig. 7, which demonstrates an instantaneous reflectivity modulation induced by IR pumping. Although the real-carrier effects can be relatively fast [≲10 ps (Refs. 12 and 15)], they occur on time scales significantly longer than those indicated in Fig. 7. We thus conclude that the data presented in Fig. 7 are consistent with the Kerr effect. Additional evidence to confirm this conclusion comes from the linear variation with the pump power shown in the inset to Fig. 7, which is consistent with the Kerr effect, but not with multiphoton absorption.[9] It is worth noting that the intensities used here are in the range 0.6–3 GW/cm$^2$, which is comparable to that used by Mondia *et al.* for GaAs,[17] and more than 30 times smaller then those used on silicon-based PCs by Tan *et al.*[9] This is in reasonable agreement with the nonlinear refractive index of $Al_xGa_{1-x}As$, which is comparable to that of GaAs, but around two orders of magnitude larger than that of silicon.

The detailed shape of the nonlinear response induced by the virtual-carrier nonlinearity is somewhat surprising. In the simplest model, we might simply expect a shift of the photonic resonance according to Eq. (1). Such a shift is indeed observed at positive time delays when free-carrier nonlinearities are dominant, but no clear shift is observed in the data at zero delay. This is in fact consistent with the experimental conditions that apply. Taking the $Al_xGa_{1-x}As$ Kerr coefficient value of $n_2 = 4 \times 10^{-13}$ W$^{-1}$ cm$^2$ for around 860 nm,[19] and the intensity we used for the experiment, we estimate a shift of −0.1 nm, which is smaller than the resolution of our experimental setup. The shape of the differential reflectivity spectra given in Fig. 6 near zero delay actually appears to imply a bleaching of the resonance occurring on a very fast time scale. This bleaching suggests that the main effect of the IR pump is to induce a change of the imaginary part of the refractive index (i.e., changing the absorption) rather than the real part. We note that the particular regime that we are considering here, namely, the Kerr nonlinearity induced by an IR pump on a photonic mode at much higher frequency, has not been investigated in detail before to our knowledge.

## V. CONCLUSIONS

In conclusion, we have demonstrated an ultrafast reflectivity modulation in $Al_xGa_{1-x}As/In_yAl_xGa_{1-x-y}As$ active photonic crystals around 800 nm by using two mechanisms, i.e., the optical Kerr effect and the ac Stark effect. Pumping in the IR below the two-photon absorption threshold was used to induce a pulse-limited reflectivity change via virtual-carrier nonlinear absorption. Additionally we demonstrated an instantaneous resonant nonlinearity using the exciton blueshift to alter the reflectivity of the photonic structure. Our results indicate that there is good potential for exploiting the virtual-carrier nonlinearities of III-V materials in ultrafast photonic switching applications, although care will need to be taken to prevent the effects from being obscured by slower real-carrier nonlinearities.

## ACKNOWLEDGMENTS

We acknowledge support from the EPSRC by Grant No. GR/S76076.